%% file: make_astro.tex
\documentclass{mpe_report}

\usepackage{psfig,graphicx,epsfig}
\usepackage{color}
\usepackage{amsmath,amssymb,epic,eepic,array}

\begin{document}

\pagenumbering{arabic}
\setcounter{page}{189}

 \renewcommand{\FirstPageOfPaper }{189}\renewcommand{\LastPageOfPaper }{192}\include{./mpe_report_petri2}             \clearpage

\end{document}

%% file: mpe_report_petri2.tex
\title{Forced oscillations in relativistic accretion disks and QPOs}
\author{J. P\'etri\inst{1}}  
\institute{Max-Planck-Institut f\"ur Kernphysik,
  Saupfercheckweg 1, 69117 Heidelberg - Germany}
\maketitle

\begin{abstract}
  In this work we explore the idea that the high frequency QPOs
  observed in LMXBs may be explained as a resonant coupling between
  the neutron star spin and epicyclic modes of accretion disk
  oscillations. We propose a new model for these QPOs based on forced
  oscillations induced in the accretion disk due to a stellar
  asymmetric rotating gravitational or magnetic field. It is shown
  that particles evolving in a rotating non-axisymmetric field are
  subject to three kinds of resonances: a corotation resonance, a
  Lindblad resonance due to a driving force, and a parametric
  resonance due to the time varying epicyclic frequencies.  These
  results are extends by means of 2D numerical simulations of a
  simplified version of the accretion disk.  The simulations are
  performed for the Newtonian gravitational potential, as well as for
  a pseudo-general relativistic potential, which enables us to explore
  the behavior of the resonances around both rotating neutron stars
  and black holes.  Density perturbations are only significant in the
  region located close to the inner edge of the disk near the ISCO
  where the gravitational or magnetic perturbation is maximal.  It is
  argued that the nearly periodic motion induced in the disk will
  produce high quality factor QPOs.
  
  Finally, applying this model to a typical neutron star, we found
  that the strongest response occurs when the frequency difference of
  the two modes equals either the spin frequency (for ``slow
  rotators'') or half of it (for ``fast rotators''). The two main
  excited modes may both be connected to vertical oscillations of the
  disk. We emphasize that strong gravity is not needed to excite the
  modes.
\end{abstract}

\section{Introduction}
\label{sec:intro}

Quasi-periodic oscillations~(QPOs) have been observed in accretion
disks around neutron stars, black holes, and white dwarf binaries with
frequencies ranging from a few~0.1~Hz up to 1300~Hz.  Recent
observations have shown a strong correlation between the low and high
frequency QPOs (Mauche~\cite{Mauche2002}, Psaltis et
al.~\cite{Psaltis1999}). This relation holds over more than 6~orders
of magnitude in frequency and strongly supports the idea that the QPO
phenomenon is a universal physical process independent of the nature
of the compact object.
  
To date, quasi-periodic oscillations (QPOs) have been observed in
about twenty Low Mass X-ray Binaries (LMXBs) containing an accreting
neutron star. Among these systems, the high-frequency QPOs (kHz-QPOs)
which mainly show up in pairs, denoted by frequencies $\nu_1$ and
$\nu_2>\nu_1$, possess strong similarities in their frequencies,
ranging from~300~Hz to about~1300~Hz, as well as in their
shapes (van der Klis~\cite{vanderKlis2000}).
 
Several models have been proposed to explain the kHz-QPOs in LMXBs. A
beat-frequency model was introduced to explain the commensurability
between the twin kHz-QPOs frequency difference and the neutron star
rotation. This interaction between the orbital motion and the star
rotation happens at some preferred radius. Alpar \& Shaham
(\cite{Alpar1985}) and Shaham~(\cite{Shaham1987}) proposed the
magnetospheric radius to be the preferred radius. The sonic-point
beat-frequency model was suggested by Miller et
al.~(\cite{Miller1998}). In this model, the preferred radius is the
point where the radial inflow becomes supersonic.

The relativistic precession model introduced by Stella \& Vietri
(\cite{Stella1998},~\cite{Stella1999}) makes use of the motion of a
single particle in the Kerr-spacetime.  In this model, the kHz-QPOs
frequency difference is related to the relativistic periastron
precession of weakly elliptic orbits while the low-frequencies QPOs
are interpreted as a consequence of the Lense-Thirring precession.
Abramowicz \& Klu{\'z}niak~(\cite{Abramowicz2001}) introduced a
resonance between orbital and epicyclic motion that can account for
the 3:2 ratio around Kerr black holes leading to an estimate of their
mass and spin. The 3:2 ratio of black hole QPOs frequencies is well
established (McClintock \& Remillard, \cite{MacClintock2003}).  In
other models, the QPOs are identified with gravity or pressure
oscillation modes in the accretion disk (Titarchuk et
al.~\cite{Titarchuk1998}, Wagoner et al.~\cite{Wagoner2001}).
Rezzolla et al. (\cite{Rezzolla2003}) suggested that the high
frequency QPOs in black hole binaries are related to p-mode
oscillations in a non Keplerian torus.

We propose a new explanation of this phenomenon based on a resonance
in the star-disk system arising from the response of the accretion
disk to either a non-axisymmetric rotating gravitational field, for a
hydrodynamical disk, or to a non-axisymmetric rotating magnetic field,
for an MHD disk.

\section{Hydrodynamical disk}

We first consider a hydrodynamical disk evolving in an asymmetric
gravitation field imposed by the rotating accreting compact object.  A
simplified linear analysis reveals the main characteristic of the
perturbed motion in the disk. This is then confirmed by full
non-linear numerical simulations.

\subsection{Linear analysis}

By perturbing the set of hydrodynamical equations governing the
evolution in the disk around the equatorial plane, and introducing the
Lagrangian displacement~$\vec{\xi}$, the weak oscillations in the
radial and vertical directions can be cast into a partial differential
equation for the radial and vertical displacement respectively.
Neglecting the sound propagation term (not important because not
leading to resonance conditions), the Laplacian displacement
$\xi_{r/z}$ satisfies~:
\begin{eqnarray}
  \label{eq:PDEXir}
  \frac{D^2\xi_\mathrm{r}}{Dt^2} & + & 
  \kappa_\mathrm{r}^2\, \xi_\mathrm{r} + \frac{1}{\rho\,r} \, 
  \frac{\partial}{\partial r}\left( r\,\rho\,\xi_\mathrm{r}\right) \, \delta g_\mathrm{r}
  = \delta g_\mathrm{r} \\
  \label{eq:PDEXiz}
  \frac{D^2\xi_\mathrm{z}}{Dt^2} & + & \kappa_\mathrm{z}^2\,\xi_\mathrm{z}
  + \frac{1}{\rho} \, \frac{\partial}{\partial z}
  \left( \rho \, \xi_\mathrm{z} \right) \, \delta g_\mathrm{z} 
  = \delta g_\mathrm{z}
\end{eqnarray}
We have introduced the radial/vertical epicyclic
frequency~$\kappa_\mathrm{r/z}$ and the perturbation in
radial/vertical gravitational field by~$\delta g_\mathrm{r/z}$. The
convective derivative is denoted by~$D/Dt = \partial_t +
\Omega\,\partial_\varphi$ where $\Omega$ is the unperturbed local
orbital frequency in the disk. The 3~terms in Eq.(\ref{eq:PDEXir}) and
(\ref{eq:PDEXiz}) are: a harmonic oscillator at
frequency~$\kappa_{r/z}$, second term on the left hand side, with a
periodically perturbed eigenfrequency, third term on the LHS, and a
periodic driven source, on the right hand side.

A careful analysis of this equation shows the emergence of three kinds
of resonance corresponding to~:
\begin{itemize}
\item a corotation resonance~;
\item an inner and outer Lindblad resonance~;
\item a parametric resonance when $m \, |\Omega_* - \Omega| = 2 \,
  \kappa_\mathrm{r/z}/n$, where $\Omega_*$ is the stellar spin,
  $\kappa_\mathrm{r/z}$ the radial and vertical epicyclic frequencies,
  $n$ is a natural integer and $m$ the azimuthal mode of the
  gravitational perturbation.
\end{itemize}

For a Newtonian disk, we have $\Omega = \kappa_r = \kappa_z$, and the
parametric resonance condition simplifies into~:
\begin{equation}
  \label{eq:ResPara}
  \frac{\Omega}{\Omega_*} = \frac{m}{m \pm 2/n}  
\end{equation}
As a consequence, the resonances are all located in the frequency
range~$\Omega\in[\Omega_*/3, 3\,\Omega_*]$. 

For the general-relativistic disk, it splits into the two following
cases~:
\begin{equation}
  \label{eq:ResParaGR}
  \Omega(r,a) \pm \frac{2\,\kappa_\mathrm{r/z}(r,a)}{m\,n} = \Omega_* 
\end{equation}
For a typical neutron star, we choose:
\begin{itemize}
\item mass~$M_*=1.4\,\mathrm{M_\odot}$~;
\item angular velocity~$\nu_*=\Omega_*/2\pi=300-600~\mathrm{Hz}$~;
\item moment of inertia~$I_*=10^{38}\;\mathrm{kg\,m^2}$~;
\item angular momentum~$a_*=c\,I_*\,\Omega_* /G\,M_*^2$.
\end{itemize}
The angular momentum is then given by~$a_*=5.79*10^{-5}\,\Omega_*$.
For the selected spin rate of the star we find~$a_*=0.109-0.218$ and
so the vertical epicyclic frequency is close to the orbital
one~$\kappa_z\approx\Omega$. Thus for the vertical resonance, we are
still close to the Newtonian case given by Eq.(\ref{eq:ResPara}).

The results for an accretion disk evolving in a Newtonian and a Kerr
spacetime are shown in Table~\ref{tab:ResPara} and \ref{tab:ResParaGR}
(P\'etri~\cite{art1}).

\begin{table}[h]
  \centering
  \begin{tabular}{| c | c c | c c |}
    \hline
     & \multicolumn{4}{c|}{Orbital frequency $\nu(r,a)$ (Hz)} \\
    \hline
    \hline
    Mode & \multicolumn{2}{c|}{$\nu_*=600$~Hz} & \multicolumn{2}{c|}{$\nu_*=300$~Hz} \\
    $m$ & $n=1$ & $n=2$ & $n=1$ & $n=2$ \\
    \hline
    \hline
    1 & -600 / 200 & ---- / 300 & -300 / 100 & --- / 150 \\
    2 & ---- / 300 & 1200 / 400 &  --- / 150 & 600 / 200 \\ 
    3 & 1800 / 360 &  900 / 450 &  900 / 180 & 450 / 225 \\
    \hline
  \end{tabular}
  \caption{Orbital frequencies at the parametric
    vertical resonance for the first three order~$n$ in the case of a
    Newtonian gravitational potential.  The value on the left of the
    symbol~/ corresponds to the absolute value sign taken to be~-
    and on the right to be~+.}
  \label{tab:ResPara}
\end{table}

\begin{table}
  \centering
  \begin{tabular}{| c | c c | c c |}
    \hline
    & \multicolumn{4}{c|}{Orbital frequency $\nu(r,a_*)$ (Hz)} \\
    \hline
    \hline
    Mode & \multicolumn{2}{c|}{$\nu_*=600$~Hz} & \multicolumn{2}{c|}{$\nu_*=300$~Hz} \\
    $m$ & $n=1$ & $n=2$ & $n=1$ & $n=2$ \\
    \hline
    \hline
    1 & ---- / 200 & ---- / 300 &  --- / 100 & --- / 150 \\
    2 & ---- / 300 & 1198 / 400 &  --- / 150 & 599 / 200 \\ 
    3 & 1790 / 360 &  899 / 450 &  898 / 180 & 450 / 225 \\
    \hline
  \end{tabular}
  \caption{Same as Tab.~\ref{tab:ResPara} but for the general-relativistic disk.}
  \label{tab:ResParaGR}       % Give a unique label
\end{table}

\subsection{Two-dimensional simulations}
    
\begin{figure}
  \centerline{\psfig{file=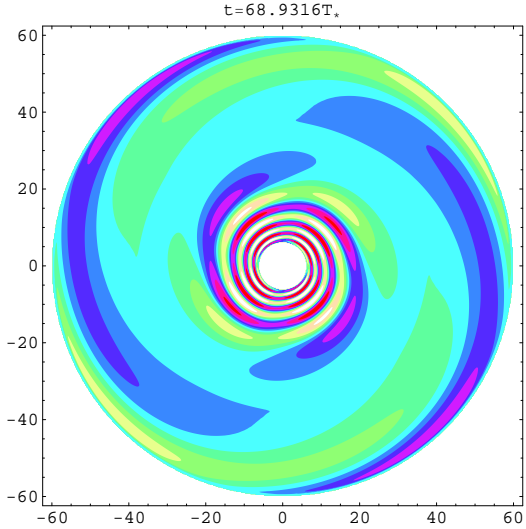,width=7cm,clip=} }
  \caption{Final snapshot of the density perturbation~$\delta\rho/\rho_0$
    in the accretion disk evolving in a quadrupolar perturbed
    Newtonian potential. Time is normalized to the spin period. The
    $m=2$ structure emerges in relation with the~$m=2$ quadrupolar
    potential perturbation.}
  \label{fig:DensDiscNewt} 
\end{figure}

\begin{figure}
  \begin{center}
    \centerline{\psfig{file=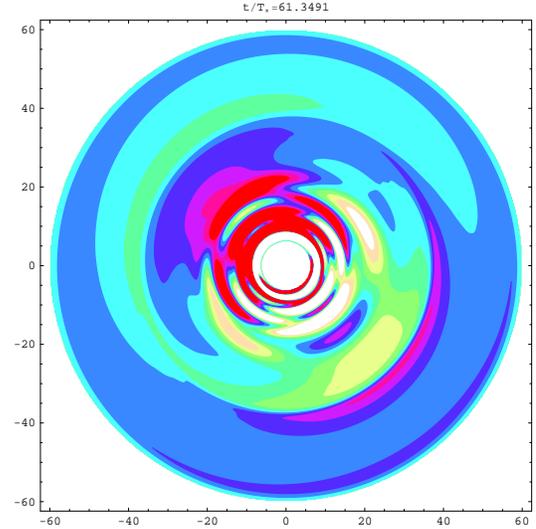,width=7cm,clip=} }
    \caption{Final snapshot of the density perturbation in the accretion disk
      evolving in a perturbed Newtonian potential having many
      azimuthal modes~$m$.}
    \label{fig:DensDiscBoiteux} 
  \end{center}
\end{figure}

From the analytical analysis of the linear response of a thin
accretion disk in the 2D limit, we know that waves are launched at the
aforementioned resonance loci. They propagate in some permitted
regions inside the disk, according to the dispersion relation obtained
by a WKB analysis.  We confirm and extend these results by performing
non linear hydrodynamical numerical simulations using a
pseudo-spectral code solving Euler's equations in a 2D cylindrical
coordinate frame.  Simulations were performed for the Newtonian as
well as for a pseudo-Newtonian potential.  For instance, in the
Newtonian potential, for a perturbation of mode~$m=2$, the stationary
density perturbation is shown in Fig.~\ref{fig:DensDiscNewt}. The
forbidden region where no wave propagation is possible, is located
between $r=24.0$ and $r=49.0$ as predicted by the WKB analysis. Outside
this region, an $m=2$ spiral structure forms and rotates at the star
speed.  The corotation resonance located at~$r=40.0$ is not seen at
this stage. When many azimuthal modes are excited at the same time as
in Fig.~\ref{fig:DensDiscBoiteux}, it becomes hard to distinguish
between the different permitted and forbidden regions for each~$m$. No
clear symmetric pattern emerges from the simulations due to
overlapping of many~$m$.  Nevertheless, in all case, Newtonian or
pseudo-Schwarzschild and pseudo-Kerr, the simulations agree well with
the detailed linear analysis.

\section{MHD disk}

We extend the previous idea to a magnetised accretion disk evolving in
the rotating asymmetric magnetic field field of the accreting compact
object (linear analysis and numerical simulations)
(P\'etri~\cite{art2}).  This case is well suited for binaries
containing a neutron star or even for white dwarfs in cataclysmic
variables for which the QPO-phenomenology seems to be identical
(Warner et al.~\cite{Warner2003}).

\subsection{Linear analysis}

Here again, applying the linear perturbation theory to the set of MHD
equations governing the motion in the disk and focusing only on
resonance terms in the radial and vertical directions, the Laplacian
displacement~$\xi_{r/z}$ satisfies~:
\begin{eqnarray}
  \label{eq:PDEXiRZ}
  \frac{D^2\xi_z}{Dt^2} + \left[ \kappa_z^2 - \frac{\partial}{\partial z}
    \left( \frac{\delta B_*^r }{\mu_0\,\rho} \, 
      \frac{\partial B_r}{\partial z} \right) \right] \, \xi_z & = & 
  \frac{\partial}{\partial z} \left( 
    \frac{B_r\,\delta B_*^r}{\mu_0\,\rho} \right) \\
  \frac{D^2\xi_r}{Dt^2} + \left[ \kappa_r^2 - \frac{\partial}{\partial r}
    \left( \frac{\delta B_*^z}{\mu_0\,\rho} \, \frac{\partial B_z}{\partial r}
    \right) \right] \, \xi_r & = &
  - \frac{\partial}{\partial r} \left( \frac{B_z\,\delta B_*^z}{\mu_0\,\rho} \right) 
\end{eqnarray}
Now, the perturbation in magnetic field~$\delta B_{r/z}$ replaces the
perturbation in gravitational field. We recognise again a harmonic
oscillator with periodically varying eigenfrequency superposed on a
driven source term. Therefore, despite the presence of a magnetic
field, the behavior of the accretion disk, at least in the linear
stage of its evolution, agrees with the study already made for a
hydrodynamical disk.  The orbital frequencies for resonance are
therefore also given by tables~\ref{tab:ResPara} and
\ref{tab:ResParaGR} for a Newtonian and a relativistic disk
respectively.

\subsection{Two-dimensional simulations}

We again performed 2D numerical simulations by solving the
magnetohydrodynamical equations for the accretion disk. This is done
by extending the previous pseudo-spectral method to a simplified
version of the 2D MHD accretion disk. 

Before the time~$t=0$, the disk stays in its axisymmetric equilibrium
state and possesses only azimuthal motion. At~$t=0$, we switch on the
perturbation by adding an asymmetric rotating component to the
magnetic field. We then let the system evolve during more than one
thousand orbital revolutions of the inner edge of the disk.

We ran a simulation in which the rotation of the star is taken into
account. This shifts the location of the ISCO closer to the surface of
the neutron star as compared to the non-rotating case.

\begin{figure}
  \begin{center}
    \centerline{\psfig{file=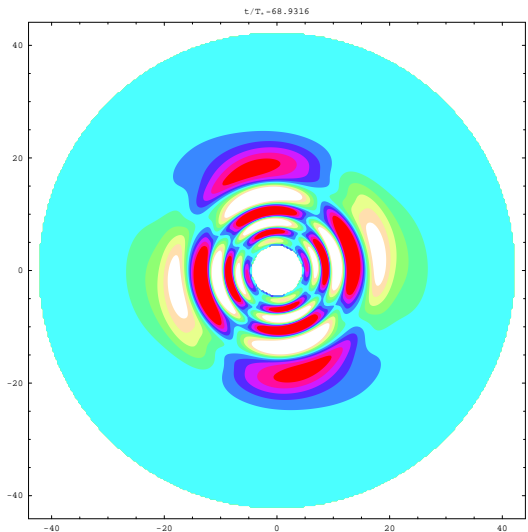,width=7cm,clip=} }
    \caption{Final snapshot of the density perturbation in the accretion disk
      evolving in a perturbed pseudo-Kerr potential with~$a=0.5$. The
      outer Lindblad resonance is not on the grid.}
    \label{fig:DensDiscKerr} 
  \end{center}
\end{figure}

We chose a star with an angular momentum of~$a_*=0.5$. Therefore, the
disk inner boundary corresponding to the marginally stable circular
orbit approaches the horizon. An example of the density perturbation
in the pseudo-Kerr metric is shown in Fig.~\ref{fig:DensDiscKerr} for
the~$m=2$ mode. The inner Lindblad radius is clearly identified while
the outer Lindblad radius is outside the simulation box.

\section{Slow vs fast rotator}

Focusing on accreting neutron stars in LMXBs, observations reveal that
they can be divided into two categories (van der
Klis~\cite{vanderKlis2004})~: the slow rotators possessing a rotation
rate $\nu_* \approx 300$ Hz, and for which the frequency difference
between the two peaks is around $\Delta \nu \approx \nu_*$ and $\nu_2
\approx 3\,\nu_1/2$ and the fast rotators having $\nu_*\approx 600$
Hz, for which this difference is around $\Delta \nu \approx \nu_*/2$.

The model presented in this work can account for this segregation if
the innermost stable circular orbit (ISCO) is taken into account.
Indeed, for a typical neutron star, the orbital frequency at the ISCO
is $\nu_\mathrm{isco} = 1571$~Hz which is therefore the upper limit
for any QPO frequency.  Discarding the resonance frequencies in the
relativistic disk which are higher than $\nu_\mathrm{isco}$, we
conclude from Table~\ref{tab:ResParaGR} that~:
\begin{itemize}
\item for slow rotators, the two highest frequencies are less than
  $\nu_\mathrm{isco}$ and given by~$\nu_1 = 599$~Hz and $\nu_2 =
  898$~Hz, therefore $\Delta\nu = 299$~Hz which is very close
  to~$\nu_*$. Moreover $\nu_2 \approx 3\,\nu_1/2$ in accordance with
  observations~;
\item for fast rotators, the highest frequency is not observed because
  the resonance is located inside the radius of the ISCO
  ($\nu_\mathrm{isco} < 1790$~Hz).  Therefore the two highest
  observable frequencies are $\nu_1 = 899$~Hz and $\nu_2 = 1198$~Hz,
  having a difference $\Delta\nu = 299$~Hz which is close
  to~$\nu_*/2$.
\end{itemize}
As already claimed in the previous section, these conclusions apply to
hydrodynamical (P\'etri~\cite{art3}) as well as to magnetised
(P\'etri~\cite{art4}) accretion disks. Observations therefore strongly
support our resonance model.

\section{Conclusion}

The consequences of a weakly rotating asymmetric stellar gravitational
or magnetic field on the evolution of a thin accretion disk are as
follows. Corotation, driven and parametric resonances are excited at
some preferred radii. The kHz-QPOs are interpreted as the orbital
frequency of the disk at locations where the vertical response to the
resonances is maximal. The 3:2 ratio is predicted for the strongest
modes and a clear distinction exists between slow and fast rotators, a
direct consequence of the presence of an ISCO. Nevertheless general
relativistic effects are not required to excite these resonances. They
behave identically in the Newtonian as well as in the Kerr field.
Therefore the QPO phenomenology is explained by the same picture,
irrespective of the nature of the compact object (black hole, neutron
star or white dwarf).  Indeed the presence or the absence of a solid
surface, a magnetic field or an event horizon plays no relevant role
in the production of the X-ray variability.

\begin{acknowledgement}
  This work was supported by a grant from the G.I.F., the
  German-Israeli Foundation for Scientific Research and Development.
  I am grateful to John Kirk for carefully reading the manuscript.
\end{acknowledgement}

% Example list of References

%\end{document}